\documentstyle[aps,epsfig]{revtex}
\begin{document}
\title{Charged and neutral hyperonic effects on the driplines}

\author{P. Roy Chowdhury$^1$\thanks{E-mail: partha$_{-}$26@hotmail.com},
C. Samanta$^{1,2}$\thanks{E-mail: csamanta@vcu.edu, chhanda.samanta@saha.ac.in} }
\address{ $^1$ Saha Institute of Nuclear Physics, 1/AF Bidhan Nagar, Kolkata 700 064, India }
\address{ $^2$ Physics Department, Virginia Commonwealth University, Richmond, VA 23284-2000, U.S.A. }

\author { D.N. Basu\thanks{E-mail: dnb@veccal.ernet.in}}
\address {Variable  Energy  Cyclotron  Centre,  1/AF Bidhan Nagar, Kolkata 700 064, India}
\date{\today}
\maketitle
\begin{abstract}
         Modification of neutron and proton driplines by the capture of strange hyperon(s) by normal nuclei has been investigated. A generalised mass formula (BWMH) based on the strangeness dependent extended liquid drop model is used to calculate the binding energy of normal nuclei as well as strange hypernuclei. The neutron ($S_n$) and proton ($S_p$) separation energies of all hypernuclei with neutral hyperons $\Lambda^0$, double $\Lambda^0$ or charged hyperons $\Xi^-$, $\Theta^+$ inside are calculated using BWMH mass formula. The normal neutron and proton driplines get modified due to the addition of the hyperon(s) ($\Lambda, ~\Lambda\Lambda, ~\Xi^-, ~\Theta^+$ etc.) to the core of normal nuclei. The hypernuclei containing the charged hyperon(s) like those with neutral hyperon(s) have similar nucleon separation energies like core nuclei if proton number instead of net charge is used in the symmetry term. Due to the effect of opposite charges present in $\Theta^+$ and $\Xi^-$ hyperons their corresponding driplines get separated from each other. All the hyperons modify mean field potential due to strong hyperon-nucleon coupling. Addition of a single charged hyperon in normal nuclei affects the entire proton drip line more prominently than that by neutral hyperon. The neutral hyperonic effect on proton dripline is significant for lighter nuclei than for heavier ones whereas both the charged as well as neutral hyperons affect almost the entire neutron dripline.

\end{abstract}
\vskip 0.2cm

\section{Introduction}
\label{section1}
Current nuclear physics is focussed on exploring fundamental interaction between nucleon and hyperon ($Y$) and basic properties of the hadronic system \cite{Keil00,Keil02}.  The worldwide activity in doing many sophisticated experiments \cite{Ba90,Ta01,Saha05,step03} in these interdisciplinary fields shows the strong scientific interest in exploring the hypernuclear properties. The nuclear dynamics has been predicted \cite{gies02} to evolve from mean-field dominance to a new kind of correlation dynamics at the neutron and proton driplines. At extreme neutron to proton ratios short-range correlations do not strictly follow the properties of single particle moving in a static mean field potential. Its effects have far reaching consequences in case of exotic nuclei as well as hypernuclear domain. The strong $Y-Y$ attractive interaction is useful to explain the phase transition in the interior of the compact stars  like strange quark stars, hybrid stars, hypercompact hyperon stars etc. composed of the strongly interacting densest matter in the universe undergoing the deconfinement phase transition\cite{hana02,mishu02}. Therefore it is interesting to study the change of dripline nuclei due to the addition of various hyperon(s) inside the core of normal nuclei which sheds some light on the interactions among hadrons and their internal quark structure in a phenomenological way. The total nuclear binding energy of any nuclei plays significant role in determining the degree of nuclear stability and hence it can be used to examine the single nucleon binding near the dripline region for all nuclei. The semi empirical mass formula proposed by Bethe-Weizs\"acker was extended for light nuclei (BWM) \cite{Sa02} and the more generalised version (BWMH) applicable for both strange and nonstrange nuclei was developed recently \cite{cs06}.
This mass formula (BWMH) based on the extended liquid drop model has been used to calculate the neutron ($S_n$) and proton ($S_p$) separation energies of both the normal and hypernuclei containing charged or neutral hyperon(s).  BWMH matches the available experimental data \cite{Ba90,Ta01} for the hyperon separation energies. Using this formula the neutron ($S_n$) and proton ($S_p$) separation energies of all normal nuclei and hypernuclei containing $\Lambda, ~\Lambda\Lambda, ~\Xi^-, \Theta^+$\cite{Na03,burk05} hyperons are calculated to predict dripline nuclei for both the normal and hypernuclei. The effect of the addition of hyperon(s) inside the core of normal nuclei has been studied near the dripline.  The effect of two oppositely charged hyperons $\Xi^-$ and $\Theta^+$ on the nuclear bindings and the driplines are demonstrated. For the hypernuclei containing positively charged hyperon like $\Theta^+$ the separation between the stability line and the proton dripline shrinks and the corresponding gap between the stability line and neutron dripline expands with respect to normal dripline. On the other hand, the proton dripline goes away with the simultaneous shifting of neutron dripline towards stability line when net charge effect is taken into account in the symmetry term for the hypernuclei containing negatively charged hyperon (e.g; $\Xi^-$). Both the neutron and proton drip lines go away from the stability line for the hypernuclei containing neutral hyperon(s) (e.g; $\Lambda$, $\Lambda\Lambda$) and the effect becomes more prominent in case of double $\Lambda$-hypernuclei. Since the mean field gets modified according to the type of hyperon inside the normal nuclear core, the values of nucleon binding (e.g; $S_p$ and $S_n$) are different for various hypernuclei.
\section{Strangeness dependent liquid drop model and nuclear stability}
\label{section2}

On the basis of the liquid drop model a generalised mass formula (BWMH) for both strange and non-strange nuclei was prescribed \cite{cs06}. The BWMH explicitly includes the strangeness and a SU(6) symmetry breaking mass term for  hyperon(s) confined inside the hypernuclei. The total binding energy of any hypernuclei of total mass number A and net charge Z containing any type of  charged or neutral hyperon(s)  is given as

\begin{eqnarray}
B(A,Z) = &&15.777A-18.34A^{2/3}-0.71\frac{Z(Z-1)}{A^{1/3}}-\frac{23.21(N-Z_c)^2}{[(1+e^{-A/17})A]}\nonumber\\
              &&+(1-e^{-A/30})\delta+ n_Y [0.0335(m_Y) - 26.7 - 48.7 \mid S \mid}{A^{-2/3}], \nonumber\\
\label{seqn1}
\end{eqnarray}
\noindent
where $\delta=12A^{-1/2}$ for $N,Z_c$ even, $=-12A^{-1/2}$ for $N,Z_c$ odd, = 0 otherwise, $n_Y$ = number of hyperons in a nucleus, $m_Y$ = mass of the hyperon in $MeV$, $S$ = strangeness of the hyperon and
mass number $A = N + Z_c + n_Y$ is equal to the total number of
baryons. $N$ and $Z_c$ are the number of neutrons and protons
respectively while the $Z$ in eqn.(1) is  given by $Z = Z_c + n_Y q$
where $q$ is the charge number (with proper sign) of
hyperon(s) constituting the hypernucleus. For non-strange (S=0)
normal nuclei, $Z_c = Z$ as $n_Y$ =0. However, as a case study, the net charge number Z has also been used instead of $Z_c$ in the symmetry term.  The choice of $\delta$
value depends on the number of neutrons and protons being odd or even  in both the cases of normal
and hypernuclei. For example, in case of $^{17}_{\Lambda}O$,
$\delta=+12A^{-1/2}$ as the (N, $Z_c$) combination is even-even,
whereas, for non-strange normal $^{17}O$ nucleus $\delta=0$ for
odd-even combination of neutrons and protons.

           Nuclear stability is controlled through the competition between the attractive nucleon-nucleon strong forces and the repulsive Coulomb forces. To study the hyperonic effect on the nuclear binding of the nuclei at the dripline it is necessary to study the single neutron $(S_n)$ and single proton separation energies $(S_p)$ for all isotopes of each element using BWMH. The comparison of the normal nuclear dripline with the hypernuclear dripline illustrates the effect of the changed nuclear dynamics at the dripline.
The separation energies ($S_n$) and ($S_p$) from the hypernuclei containing any hyperon ( $\Lambda,\Lambda\Lambda, \Xi^-, \Theta^+ etc.$) inside the nucleus are defined as   
\begin{equation}
S_n = B(A,Z)_{hyper} - B(A-1, Z)_{hyper}, ~S_p = B(A,Z)_{hyper} - B(A-1, Z-1)_{hyper}.
\label{seqn2}
\end{equation}
\noindent
where $B(A,Z)_{hyper}$ is the binding energy of a hypernucleus with a hyperon(s) inside with A and Z being the total mass number and net charge number respectively. Their respective binding energies provide necessary guidelines for studying the nuclear stability near the driplines against decay by emission of protons or neutrons. The neutron dripline is defined as the last bound neutron rich nuclei beyond which the neutron separation energy changes sign and becomes negative. Additions of one neutron to the neutron drip nuclei makes the system unbound with negative single neutron separation energy (Figures 1-3). Similarly the proton dripline nucleus is defined as the last bound neutron deficient nucleus beyond which the proton separation energy becomes negative.

\section{Neutral and charged hyperonic effects on the dripline nuclei}
\label{section3}
                     Starting from dripline for normal nuclei total seven driplines are listed (Tables I - VII) and to demonstrate the charged hyperonic effect on the driplines table VIII is included. It is found that due to opposite charges in the two different hyperons ($\Xi^-,\Theta^+$) proton driplines for $\Theta^+$-hypernuclei comes closer to stability line  in most of the places than that for $\Xi^-$-hypernuclei. Since the net charge (Z) of the hypernuclei increases due to positive charge of  $\Theta^+$-hyperon, the neutron dripline goes away from the stability line in some places in comparison to the neutron dripline  of $\Xi^-$-hypernuclei. The whole calculation for table VIII has been done using eqn.(2) keeping proton number in symmetry term.

                    From the table II containing the driplines of $\Lambda$-hypernuclei, it is seen that $\Lambda$-hyperonic \cite{Huan01} effects are more prominent on the neutron dripline in comparison to respective proton dripline. The proton dripline of single $\Lambda$-hypernuclei are shifted only for few light nuclei (e.g; Li, N, O, Mg, As, Kr and Nb.) and as the number of nucleon increases the dripline merges to normal proton dripline due to the weak $\Lambda$-nucleon \cite{Taka02} coupling effect on static mean field potential. Some normal neutron drip nuclei like $Z_c=10, N=22$, $ Z_c=61, N=146$ change to $Z_c=10, N=24$ and$Z_c=61, N=148$ respectively due to the addition of single $\Lambda$ to the core so that the neutron dripline of $\Lambda$-hypernuclei goes more away from the stability line. Similar picture emerges from table III for the proton and neutron driplines of $\Lambda\Lambda$-hypernuclei. In this case relatively larger effect on proton and neutron dripline nuclei occurs  due to the enhanced hyperon-nucleon coupling effect. The proton drip line for $\Lambda\Lambda$-hypernuclei moves away from the stability line for the light and medium heavy nuclei (e.g; $Z_c$=4 to 8, 10, 12, 18, 20, 22, 24, 26, 27, 28, 33, 36, 41, 42, 52, 71 etc.). The neutron dripline of double $\Lambda$-hypernuclei are shifted more prominently for almost all nuclei  from light to heavy in comparison to the normal neutron dripline. For example, the normal neutron drip nuclei $Z_c=8, N=18$ and $Z_c=70, N=170$ get shifted to more neutron rich nuclei $Z_c=8, N=20$ and $Z_c=70, N=172$ respectively for the corresponding $\Lambda\Lambda$-hypernuclei. Increase of the number of hyperons inside the core of normal nuclei affects the nuclear dynamics to deviate from the mean field potential at the driplines and hence additional bound  nuclei are possible. Addition of neutral hyperon(s) altering the nucleon bindings of the baryonic system are illustrated in fig. 1(b, c).

Charged hyperons affect the driplines in an interesting way. For example for Holmium (proton number= 67) the proton dripline have N=78 for $\Xi^-$-hypernuclei but N=80 for
$\Theta^+$-hypernuclei which is less neutron deficient than the former. On the other hand, for same element the neutron dripline nuclei have N=162 for $\Xi^-$-hypernuclei but N=164 for $\Theta^+$-hypernuclei which contains more number of neutrons than $\Xi^-$-hypernuclei. The corresponding normal proton and neutron drip nuclei for Holmium are at N=79 and N=162 respectively. It is noteworthy that the proton-drip  $\Xi^-$ hypernucleus with proton number Zc= 67 and net charge Z=66 has N=78 and this neutron number is  reached by Z=70 normal nuclei, not Z=66. On the other hand, for the same element (proton number = 67, net charge Z=66) the neutron-drip $\Xi^-$ hypernucleus has N=162 which is same as the limiting neutron number of the Z=67 normal nucleus, not Z=66. But, the N=164 limiting number for $\Theta^+$-hypernuclei (proton number=67, net charge=68 ) is reached by Z=68 normal nucleus. In general since $\Xi^-$ is a negatively charged hyperon it neutralises one proton inside the normal core and the net positive charge of the $\Xi^-$-hypernuclei is reduced. Therefore, one expects that less number of neutrons would be needed to overcome the Coulomb repulsion. But, this is not the actual situation as strangeness and mass in addition to the net-charge play important role in determining the binding energy. Similarly, $\Theta^+$ acts as a heavy proton inside the normal core so that the effective positive charge of $\Theta^+$-hypernuclei gets increased and comparatively more number of neutrons are expected to make the baryons bound, a phenomenon not always supported by the actual situation. Thus, not only the drip-line positions, but also their detailed structure differs from the driplines of normal nuclei as the net charge as well as the mass and the strangeness of the hypernuclei produces significant effects on their binding energies.

              A charged liquid drop \cite{Myers69} consisting of fermions (like nucleons and hyperons) can be considered as a weakly interacting Fermi gas due to the large mean free path (almost equals nuclear radius) of a nucleon moving inside the nucleus and occupation of the single particle orbitals in the a simple mean field potential are restricted by the Pauli blocking \cite{Hyde}. The binding energy of the baryonic system will be maximum when nucleons and hyperons occupy the lowest possible orbitals favouring equal number of charged and neutral baryons. For hypernuclei containing charged hyperon(s) net charge (Z) instead of proton number $Z_c$ should be equal to neutron number (i.e; Z=N=A/2) to get the maximum stability from the symmetric distribution. Any other asymmetric distribution like $Z=A/2-\eta$, $N=A/2+\eta$ will lift the Fermions from the occupied into empty orbitals giving rise to the loss of symmetry energy. Thus addition of charged hyperon modify the occupancy of twofold spin-degenerate proton energy levels and changes the contribution of symmetry energy to the net nuclear binding significantly.

               The symmetry term \cite{Stein05} plays very significant role in determining the nucleon binding of the charged hypernuclei. The effect of using proton number and net charge on the dripline and the nucleon binding of $\Xi^-$ and $\Theta^+$ hypernuclei are demonstrated in the tables IV, V, VI, VII and figures 2(a,b), 3(a,b) respectively. The $S_n$ vs. N plot of hypernuclei follow the core nuclei if we take proton number in symmetry term whereas it follows the nucleon binding pattern of net charge element if net charge is used in symmetry term. In fig.3(c) it is shown that ($Li$ core + $\Theta^+$)-hypernuclei behaves as normal $Be$ element using net charge in symmetry term whereas it behaves like its core element $(Li)$ if only proton number is there in the symmetry term of BWMH.

 Binding energy of a nucleus is known to depend on the neutron-proton asymmetry. Presence of a charged hyperon inside a normal nucleus may alter the symmetry energy. To investigate its possible consequences, proton number of the symmetry term is replaced by  the net charge of the respective hypernuclei and the resultant effects on the dripline nuclei are tabulated in table V and VII. In case of negatively charged hyperon (e.g; $\Xi^-$) inside the hypernuclei, net charge in symmetry term of BWMH makes the proton dripline move further away at the expense of bringing the neutron dripline closer to the stability line. The neutron dripline  comes closer towards the $\beta$-stability line particularly for medium heavy to heavy nuclei by two nucleon shift relative to normal dripline whereas corresponding proton dripline goes away from it appreciably by one nucleon shift for light hypernuclei and two or three nucleon shift for heavier hypernuclei. For example one can compare between Table I, Table IV and TableV to see that the use of the net charge (Z) in the symmetry term in calculation of the binding energy of the $\Xi^-$-hypernuclei instead of the usual proton charge number ($Z_c$). It is seen that the neutron dripline shifts towards the $\beta$-stability line from $Z_c$=55, N=132 (normal)  $Z_c$=55, N=132($Z_c$ in symmetry term) to $Z_c$=55, N=130 (Z in the symmetry term). On the neutron deficient side,the proton drip line moves from $Z_c$=55, N=61 (normal), $Z_c$=55, N=60 ( $Z_c$ in the symmetry term) to  $Z_c$=55, N=59 (Z in the symmetry term), i.e., away from the stability line.

              Similarly in case of positively charged hypernuclei (e.g; $\Theta^+$-hypernuclei) the effect of using net charge in symmetry is just reverse.  Consideration of the net charge in the symmetry term makes the proton binding less than the normal nuclei. This effect is more prominent in the  heavier nuclei and the deviation from normal proton dripline becomes more distinct in the heavy element regions. On the contrary, the entire neutron dripline nuclei encounter two or four nucleons shifts and move the neutron dripline away from the stability line.

One can see from tableI and tableVII that addition of a $\Theta^+$ hyperon shifts the neutron-drip line away from the stability line and moves the proton drip line closer to the line of stability. For example,  normal neutron-drip nuclei are (i)$Z_c$=5, N=12; (ii)$Z_c$=78, N=190 while the neutron drip nuclei for   $\Theta^+$- hypernuclei are (i)$Z_c$=5, N=14; (ii)$Z_c$=78, N=194. The proton dripline comes from (i)$Z_c$=5, N=3(normal); (ii)$Z_c$=78, N=91 (normal) to (i)$Z_c$=5, N=4 (Z in symmetry term); (ii)$Z_c$=78, N=93 (Z in symmetry term) respectively for $\Theta^+$-hypernuclei. This is exactly opposite to the trend found in $\Xi^-$-hypernuclei.

\section{Summary and Conclusion}
\label{section4}

In summary, a generalised strangeness dependent liquid drop model (BWMH) is used to calculate the nucleonic binding of both normal and strange ($\Lambda, ~\Lambda\Lambda, ~\Xi^-, ~\Theta^+$) hypernuclei. Since the character of the mean field potential changes at the region of extreme neutron to proton ratios, it is important to study the effect of hyperons captured inside the normal core both near the stability line as well as near the driplines.  A thorough study of the effects of charged and neutral hyperons on the dripline nuclei has been carried out. It is found that the charged hyperons shift the neutron and proton driplines away from the stability line both in one direction depending on the hyperon charge. The neutral hyperons ($\Lambda, ~\Lambda\Lambda$) shift  their neutron and proton drip lines away from the stability line making the region of bound nuclei wider. The proton drip lines are however not much affected, except for few light nuclei by the addition of  single $\Lambda$. For $\Lambda\Lambda$ nuclei also proton drip line is affected mainly for light and medium heavy nuclei where it moves away from the stability line. Interestingly,  in strange hypernuclei, not only the drip-line positions, but also their detailed structure differs from the driplines of normal nuclei. This happens as the net charge in addition to the mass and the strangeness of the hypernuclei produces significant effects on their binding energies.

Addition of a $\Theta^+$ hyperon in a normal nucleus changes the net charge of the nucleus by +1. Consideration of the net charge in the symmetry term of the binding energy formula (BWMH) makes the entire proton dripline of $\Theta^+$-hypernuclei significantly shift closer to the stability line. On the contrary, the entire neutron dripline nuclei encounter two or four nucleons shifts leading the neutron dripline away from the stability line. The opposite happens for the $\Xi^-$-hypernuclei as it effectively neutralizes the charge of a proton instead of adding to it. The $S_n$ vs. N plot of hypernuclei follow the core nuclei if we take proton number in symmetry term whereas, it follows the nucleon binding pattern of net charge element if net charge (instead of the proton number) is used in the symmetry term.  As the charged hyperons modify the proton energy levels of a weakly interacting Fermi gas, the symmetry energy plays important role in determining the binding of nucleons at the driplines. This is probably the reason that the reverse effects of oppositely charged hyperons on shifting the dripline becomes more prominent if effective charge instead of proton number is used in the symmetry energy.

The modification of dripline thus points at the possible evolution of nuclear dynamics from static mean field dominance to a bound quantum system which can be described as a nucleon-nucleon mean field with additional strangeness- and hyperon mass dependence arising from the hyperon-nucleon interaction. More sophisticated experiments using radioactive ion beam (RIB) and theoretical studies on microscopic footing are necessary to explore the normal as well as these strange hypernuclei to reveal the binding and structure of these weakly bound systems near the neutron and proton driplines.         \\
      
\begin{figure}[h]
\eject\centerline{\epsfig{file=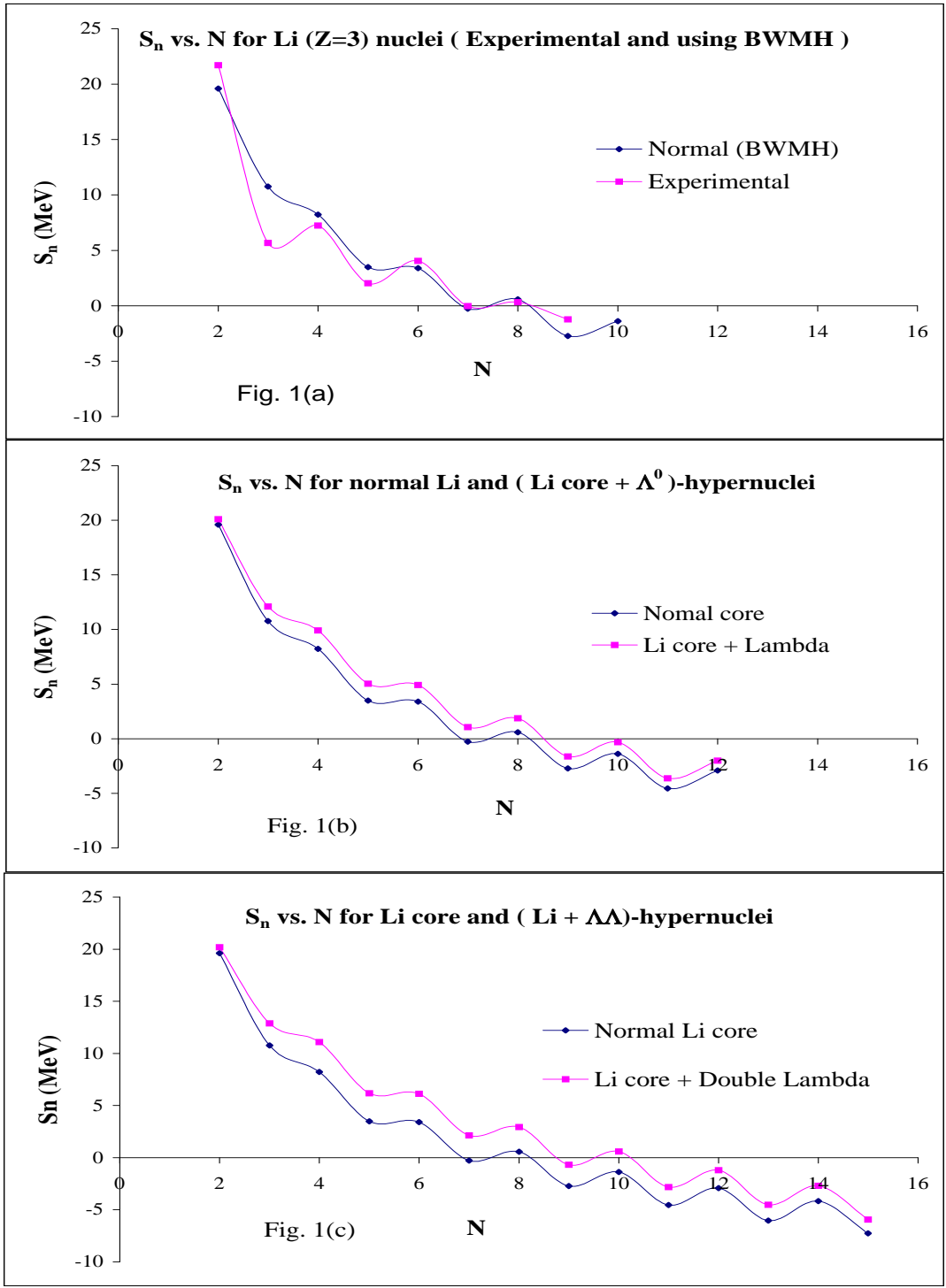,height=20cm,width=15cm, angle=0}}
\vskip 1.0cm
\caption
{ The Sn vs. N plot of Li shows the comparison of neutron binding between (a) Experimental data and BWMH calculation for normal Li nuclei, (b) Normal Li and (Li+$\Lambda$)-hypernuclei using BWMH, (c) Normal Li and (Li+$\Lambda\Lambda$)-hypernuclei using BWMH.}
\end{figure}

\begin{figure}[h]
\eject\centerline{\epsfig{file=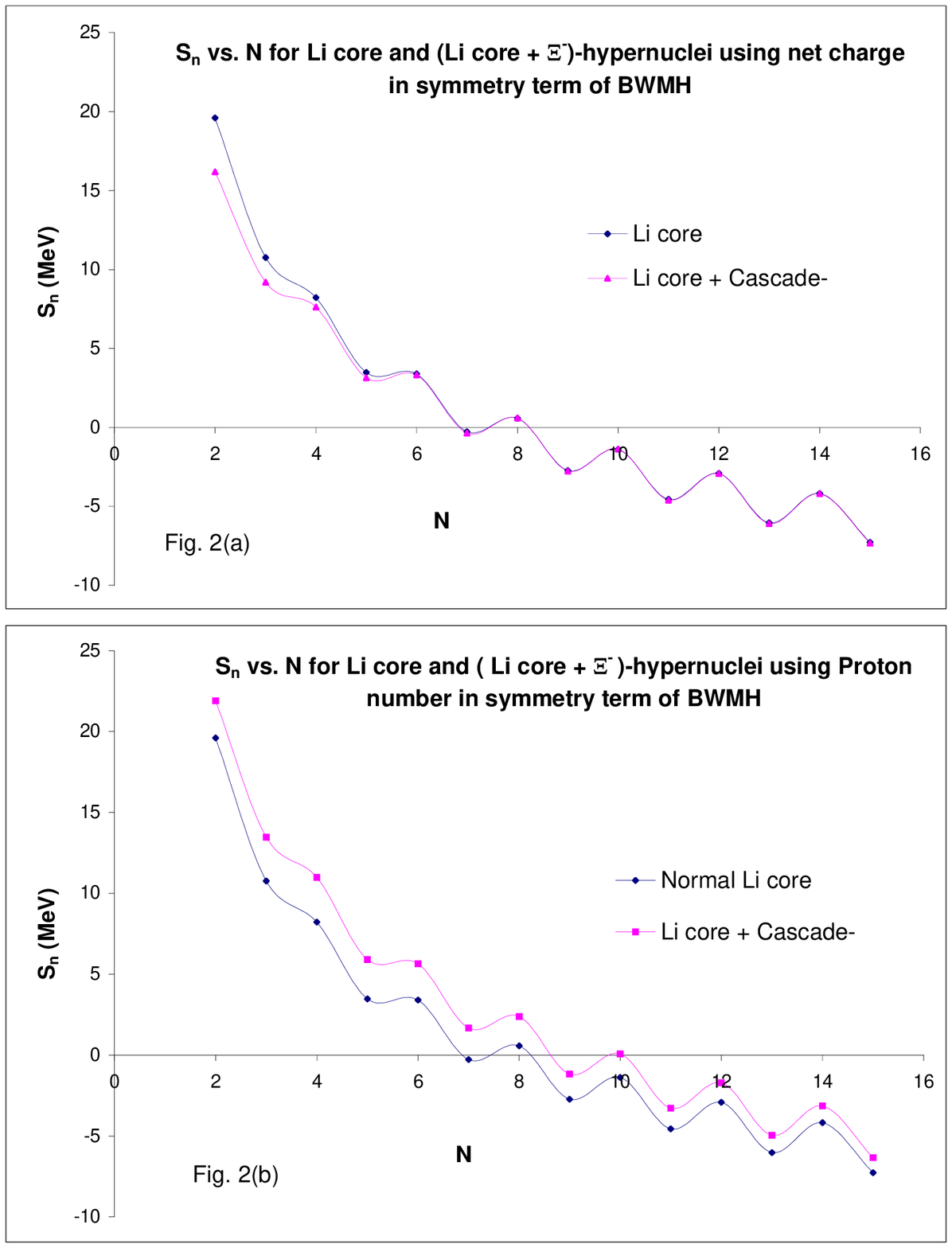,height=20cm,width=15cm, angle=0}}
\vskip 1.0cm
\caption
{  The Sn vs. N plot of Li shows the comparison of neutron binding between (a) Normal Li and (Li+$\Xi^-$)-hypernuclei using net charge in symmetry term of BWMH (b) Normal Li and (Li+$\Xi^-$)-hypernuclei using Proton number in symmetry term of BWMH }
\end{figure}

\begin{figure}[h]
\eject\centerline{\epsfig{file=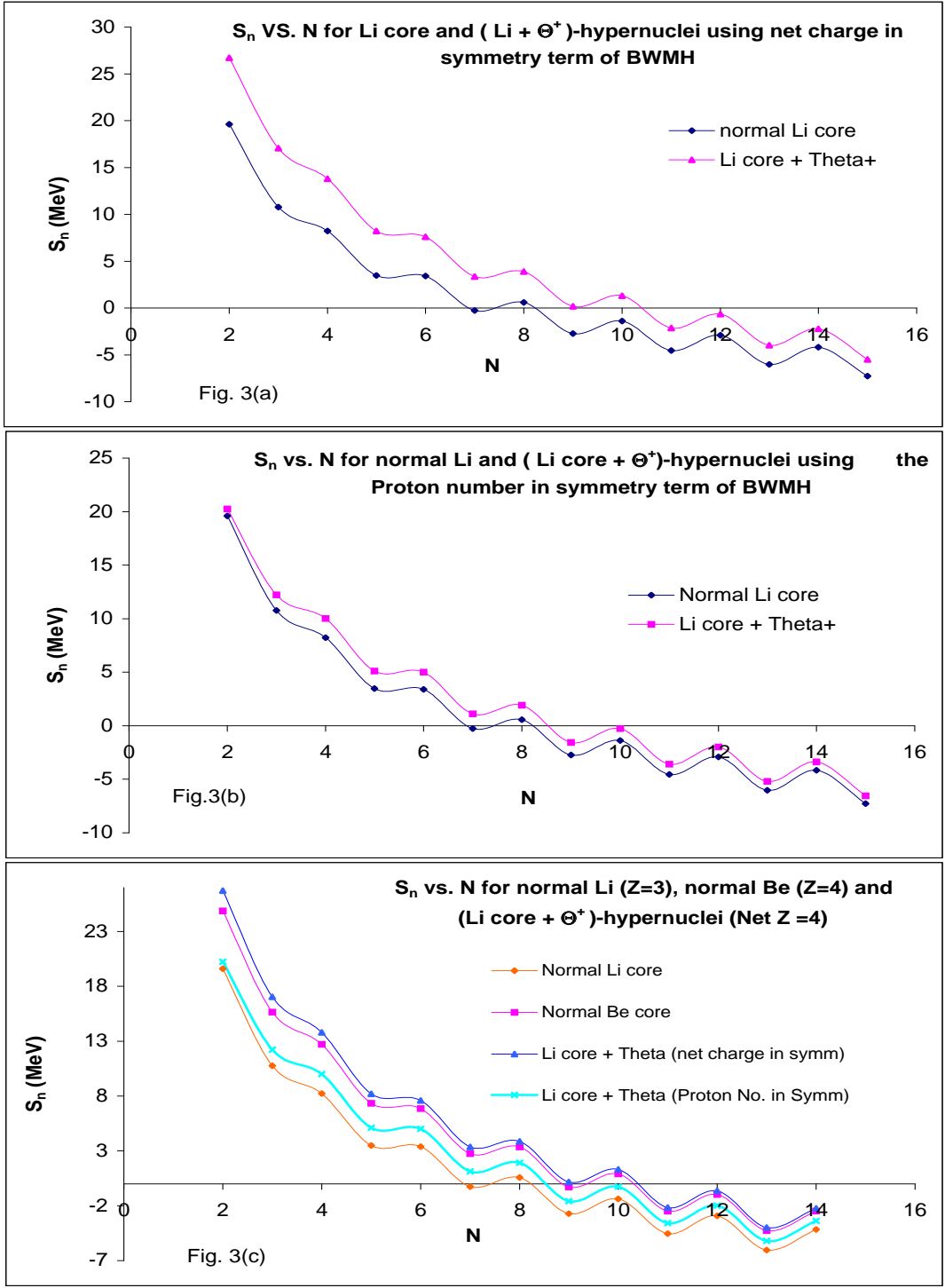,height=20cm,width=15cm, angle=0}}
\vskip 1.0cm
\caption
{ The Sn vs. N plot of Li shows the comparison of neutron binding between (a) Normal Li and (Li+$\Theta^+$)-hypernuclei using net charge in symmetry term of BWMH (b) Normal Li and (Li+$\Theta^+$)-hypernuclei using Proton number in symmetry term of BWMH. Figure 3(c) shows that Sn vs. N plot of (Li + $\Theta^+$)-hypernuclei follow (i) the Li core when proton number (=3) is used in symmetry and (ii) Be core when net charge (=4) is used in symmetry. }
\end{figure}

\begin{table}
\caption{One neutron and one proton separation energies (in MeV) on and just beyond the driplines for normal [19]  nuclei using BWM.}

\begin{tabular}[pos]{cccccccccccc}
Proton& & &One beyond&&&Neutron&&&One beyond& \\ 
drip &$S_p$&$S_n$&p-drip&$S_p$&$S_n$&drip&$S_p$&$S_n$&n-drip&$S_p$&$S_n$ \\
\hline
$Z_c, N$&$MeV$&$MeV$&$Z_c, N$&$MeV$&$MeV$&$Z_c, N$&$MeV$&$MeV$&$Z_c, N$&$MeV$&$MeV$ \\
\hline
&&&&&&&&&\\
   2,  1&  .210E+01&  .267E+02&  2,  0& -.820E+01&  .888E+02&  2,  4&  .238E+02&  .336E+01&  2,  5&  .278E+02& -.488E+00\\
  3,  2&  .336E+01&  .196E+02&  3,  1& -.349E+01&  .321E+02&  3,  8&  .270E+02&  .585E+00&  3,  9&  .292E+02& -.272E+01\\
  4,  2&  .111E+01&  .249E+02&  4,  1& -.416E+01&  .359E+02&  4, 10&  .283E+02&  .905E+00&  4, 11&  .304E+02& -.250E+01\\
  5,  3&  .741E+00&  .197E+02&  5,  2& -.326E+01&  .289E+02&  5, 12&  .255E+02&  .985E+00&  5, 13&  .273E+02& -.247E+01\\
\end{tabular}
\end{table}

\nopagebreak
\begin{table}
\caption{Proton and neutron separation energies of $\Lambda^0$-hypernuclei on and just beyond the driplines using BWMH.}

\begin{tabular}[pos]{cccccccccccc}

Proton& & &One beyond&&&Neutron&&&One beyond& \\ 
drip &$S_p$&$S_n$&p-drip&$S_p$&$S_n$&drip&$S_p$&$S_n$&n-drip&$S_p$&$S_n$ \\
\hline
$Z_c, N$&$MeV$&$MeV$&$Z_c, N$&$MeV$&$MeV$&$Z_c, N$&$MeV$&$MeV$&$Z_c, N$&$MeV$&$MeV$ \\
\hline
&&&&&&&&&\\

  Not found& --& --&  --& --& --&    2,  6&  .297E+02&  .187E+01&  2,  7&  .323E+02& -.159E+01\\
  3,  1&  .122E+01&  .289E+02&  3,  0& -.289E+01&  .509E+02&  3,  8&  .261E+02&  .187E+01&  3,  9&  .282E+02& -.162E+01\\
  4,  2&  .379E+01&  .243E+02&  4,  1& -.430E+00&  .323E+02&  4, 10&  .277E+02&  .184E+01&  4, 11&  .297E+02& -.167E+01\\
  5,  3&  .239E+01&  .195E+02&  5,  2& -.102E+01&  .278E+02&  5, 12&  .249E+02&  .173E+01&  5, 13&  .267E+02& -.179E+01\\
\end{tabular}
\end{table}

\nopagebreak
\begin{table}
\caption{Proton and neutron separation energies of $\Lambda\Lambda$-hypernuclei on and just beyond the driplines using BWMH.}

\begin{tabular}[pos]{cccccccccccc}

Proton& & &One beyond&&&Neutron&&&One beyond& \\ 
drip &$S_p$&$S_n$&p-drip&$S_p$&$S_n$&drip&$S_p$&$S_n$&n-drip&$S_p$&$S_n$ \\
\hline
$Z_c, N$&$MeV$&$MeV$&$Z_c, N$&$MeV$&$MeV$&$Z_c, N$&$MeV$&$MeV$&$Z_c, N$&$MeV$&$MeV$ \\
\hline
&&&&&&&&&\\
   Not found&  --&  --&  --&--&  --&  2,  8&  .332E+02&  .635E+00&  2,  9&  .352E+02& -.273E+01\\
   Not found&  --&  --&  --&--&  --&  3, 10&  .293E+02&  .603E+00&  3, 11&  .311E+02& -.281E+01\\
  4,  1&  .202E+01&  .299E+02&  4,  0& -.106E+01&  .440E+02&  4, 12&  .309E+02&  .593E+00&  4, 13&  .325E+02& -.285E+01\\
  5,  2&  .600E+00&  .269E+02&  5,  1& -.247E+01&  .324E+02&  5, 14&  .278E+02&  .531E+00&  5, 15&  .293E+02& -.292E+01\\
\end{tabular}
\end{table}

\nopagebreak
\begin{table}
\caption{Proton and neutron separation energies of $\Xi^-$-hypernuclei on and just beyond the driplines using net charge in Coulomb term but Proton number in asymmetry term of BWMH.}

\begin{tabular}[pos]{cccccccccccc}

Proton& & &One beyond&&&Neutron&&&One beyond& \\ 
drip &$S_p$&$S_n$&p-drip&$S_p$&$S_n$&drip&$S_p$&$S_n$&n-drip&$S_p$&$S_n$ \\
\hline
$Z_c, N$&$MeV$&$MeV$&$Z_c, N$&$MeV$&$MeV$&$Z_c, N$&$MeV$&$MeV$&$Z_c, N$&$MeV$&$MeV$ \\
\hline
&&&&&&&&&\\
   Not found&  --&--&--&--&--&  2,  8&  .359E+02&  .466E-01&  2,  9&  .378E+02& -.324E+01\\
   Not found&  --&--&--&--&--&  3, 10&  .312E+02&  .777E-01&  3, 11&  .330E+02& -.328E+01\\
  4,  1&  .216E+01&  .341E+02&  4,  0& -.455E+00&  .532E+02&  4, 12&  .324E+02&  .126E+00&  4, 13&  .341E+02& -.328E+01\\
  5,  2&  .724E+00&  .288E+02&  5,  1& -.241E+01&  .363E+02&  5, 14&  .291E+02&  .108E+00&  5, 15&  .306E+02& -.332E+01\\
\end{tabular}
\end{table}

\nopagebreak
\begin{table}
\caption{Proton and neutron separation energies of $\Xi^-$-hypernuclei on and just beyond the driplines using net charge in Coulomb term and asymmetry term of BWMH.}

\begin{tabular}[pos]{cccccccccccc}

Proton& & &One beyond&&&Neutron&&&One beyond& \\ 
drip &$S_p$&$S_n$&p-drip&$S_p$&$S_n$&drip&$S_p$&$S_n$&n-drip&$S_p$&$S_n$ \\
\hline
$Z_c, N$&$MeV$&$MeV$&$Z_c, N$&$MeV$&$MeV$&$Z_c, N$&$MeV$&$MeV$&$Z_c, N$&$MeV$&$MeV$ \\
\hline
&&&&&&&&&\\
  ~Not found& --&  --&  --& --& --&  2,  6&  .364E+02&  .657E+00&  2,  7&  .384E+02& -.265E+01\\
  ~Not found& --&  --&  --& --& --&  3,  8&  .310E+02&  .604E+00&  3,  9&  .329E+02& -.276E+01\\
  ~Not found&  --&  --&  --&--&  --&  4, 10&  .317E+02&  .627E+00&  4, 11&  .335E+02& -.279E+01\\
  5,  1&  .188E+00&  .298E+02&  5,  0& -.278E+01&  .439E+02&  5, 12&  .283E+02&  .576E+00&  5, 13&  .300E+02& -.287E+01\\
\end{tabular}
\end{table}

\nopagebreak
\begin{table}
\caption{Proton and neutron separation energies of $\Theta^+$-hypernuclei on and just beyond the driplines using net charge in Coulomb term but Proton number in asymmetry term of BWMH.}

\begin{tabular}[pos]{cccccccccccc}

Proton& & &One beyond&&&Neutron&&&One beyond& \\ 
drip &$S_p$&$S_n$&p-drip&$S_p$&$S_n$&drip&$S_p$&$S_n$&n-drip&$S_p$&$S_n$ \\
\hline
$Z_c, N$&$MeV$&$MeV$&$Z_c, N$&$MeV$&$MeV$&$Z_c, N$&$MeV$&$MeV$&$Z_c, N$&$MeV$&$MeV$ \\
\hline
&&&&&&&&&\\
  Not found&  --&--&--&--&--&        2,  6&  .291E+02&  .193E+01&  2,  7&  .316E+02& -.154E+01\\
  3,  1&  .515E+00&  .291E+02&  3,  0& -.361E+01&  .512E+02&  3,  8&  .255E+02&  .192E+01&  3,  9&  .276E+02& -.157E+01\\
  4,  2&  .317E+01&  .245E+02&  4,  1& -.106E+01&  .325E+02&  4, 10&  .272E+02&  .190E+01&  4, 11&  .292E+02& -.162E+01\\
  5,  3&  .182E+01&  .197E+02&  5,  2& -.160E+01&  .279E+02&  5, 12&  .244E+02&  .178E+01&  5, 13&  .262E+02& -.174E+01\\
\end{tabular}
\end{table}

\begin{table}
\caption{Proton and neutron separation energies of $\Theta^+$-hypernuclei on and just beyond the driplines using net charge both in Coulomb term and asymmetry term of BWMH.}

\begin{tabular}[pos]{cccccccccccc}

Proton& & &One beyond&&&Neutron&&&One beyond& \\ 
drip &$S_p$&$S_n$&p-drip&$S_p$&$S_n$&drip&$S_p$&$S_n$&n-drip&$S_p$&$S_n$ \\
\hline
$Z_c, N$&$MeV$&$MeV$&$Z_c, N$&$MeV$&$MeV$&$Z_c, N$&$MeV$&$MeV$&$Z_c, N$&$MeV$&$MeV$ \\
\hline
&&&&&&&&&\\
    2,  1&  .200E+01&  .363E+02&  2,  0& -.465E+00&  .755E+02&  2,  8&  .298E+02&  .127E+01&  2,  9&  .320E+02& -.214E+01\\
  3,  2&  .130E+01&  .267E+02&  3,  1& -.306E+01&  .385E+02&  3, 10&  .264E+02&  .130E+01&  3, 11&  .283E+02& -.215E+01\\
  4,  2&  .106E-01&  .304E+02&  4,  1& -.367E+01&  .406E+02&  4, 12&  .283E+02&  .131E+01&  4, 13&  .301E+02& -.218E+01\\
  5,  4&  .232E+01&  .208E+02&  5,  3& -.937E+00&  .240E+02&  5, 14&  .254E+02&  .123E+01&  5, 15&  .271E+02& -.226E+01\\
\end{tabular}
\nopagebreak
\end{table}
\nopagebreak
\begin{table}
\caption{Effects of negatively charged $\Xi^-$ and positively charged $\Theta^+$ hyperon on the dripline nuclei are shown in the following table using the proton number in the symmetry term of BWMH. One-nucleon separation energies (in MeV) are listed on driplines for each element with the lowest and highest number of bound neutrons in $\Xi^-$ and $\Theta^+$-hypernuclei.}

\begin{tabular}[pos]{cccccccccc}

At.&$\Xi^-$ &$\Xi^-$ &$\Theta^+$-Hyper&$\Theta^+$-Hyper&At.&$\Xi^-$&$\Xi^-$ &$\Theta^+$-Hyper&$\Theta^+$-Hyper \\
No.&p-drip&n-drip&p-drip&n-drip&No.&p-drip&n-drip&p-drip&n-drip \\
&&&&&&&&&\\
\hline
&&&&&&&&&\\
Z&$N,S_p$&$N,S_n$&$N,S_p$&$N,S_n$&Z&$N,S_p$&$N,S_n$&$N,S_p$&$N,S_n$ \\
\hline
&&&&&&&&&\\
    4&  1,   2.16& 12,    .13&  2,   3.17& 10,   1.90&  5&  2,    .72& 14,    .11&  3,   1.82& 12,   1.78\\
  6&  2,    .20& 16,    .10&  3,   1.26& 16,    .01&  7&  4,   1.33& 18,    .08&  5,   2.36& 18,    .02\\
\end{tabular}
\end{table}


\begin{references}
\bibitem{Keil00} C. M. Keil, F. Hofmann and H. Lenske 2000 Phys. Rev. C 61, 064309
\bibitem{Keil02} C. Keil, H. Lenske and C. Greiner 2002 J. Phys. G: Nucl. Part. Phys. 28, 1683
\bibitem{Ba90} H. Bando, T. Motoba, J. Zofka 1990 Int. J. Mod. Phys. A 5, 4021; see references therein.
\bibitem{Ta01} H.Takahashi et al. 2001 Phys. Rev. Lett. 87, 212502 .
\bibitem{Saha05} P.K.Saha et al. 2005 Phys. Rev. Lett. 94, 052502 . 
\bibitem{step03} S. Stepanyan et al. (CLAS) 2003 Phys. Rev. Lett. 91, 25001.
\bibitem{gies02}Annual Report 2002 Institute for Theoretical Physics I : Justus-Liebig-Universit\"at-Giessen 
\bibitem{hana02}Proceedings of the Workshop: "XEUS-studying the evolution of the hot universe" held in Garching, 2002
eds. G. Hasinger et al; MPE Report 281
\bibitem{mishu02}I. N. Mishutin et al; 2002 arxiv:hep-ph/0210422 
\bibitem{Sa02} C. Samanta and S. Adhikari 2002 Phys. Rev. C 65, 037301; ~2004 Phys. Rev. C 69, 049804; ~2004 Nucl. Phys. A 738, 491.
\bibitem{cs06}C. Samanta, P. Roy Chowdhury and D.N. Basu 2006 J. Phys. G: Nucl. Part. Phys. 32, 363. 
\bibitem{Na03} T. Nakano et al. 2003  Phys. Rev. Lett. 91, 012002.
\bibitem{burk05} Volker D. Burkert (Jefferson lab.), 2005 arXiv: hep-ph/0510309.
\bibitem{Huan01}JIA Huan-Yu, SUN Bao-Xi, MENG Jie, ZHAO En-Guang 2001 Chin.Phys.Lett. 18, 1571. 
\bibitem{Taka02} T. Takatsuka, S. Nishizaki, and Y. Yamamoto 2002 Eur. Phys. J. A 13, 213
\bibitem{Myers69}W.D. Myers, W.J. Swiatecki, Ann. Phys. 55 (1969) 395.
\bibitem{Hyde} K. Heyde, Basic Ideas and Concepts in Nuclear Physics, IOP, Bristol, 1999.
\bibitem{Stein05} A.W. Steiner, M. Prakash, J.M. Lattimer, P.J. Ellis 2005 Phys. Reports 411, 325-375.
\bibitem{shift} P. Roy Chowdhury, C. Samanta, D.N. Basu  2005  Mod. Phys. Lett. A 20, 1605.
 
\end{references}
\end{document}